\documentclass[11pt]{article}

\usepackage{amsmath}
\usepackage{amssymb}
\usepackage{extarrows}
\usepackage{graphicx}
 \usepackage{indentfirst}
\usepackage{hyperref}

\renewcommand{\baselinestretch}{1.3}
  \renewcommand{\arraystretch}{1.1}
\begin{document}

 \title{A Note On One Realization of a Scalable Shor Algorithm}

\author{Zhengjun Cao$^{1}$, \qquad Lihua Liu$^{2}$}

  \footnotetext{\noindent $^1$Department of Mathematics, Shanghai University, Shanghai,
  China. \, \textsf{caozhj@shu.edu.cn} \\
     $^2${Department of Mathematics, Shanghai Maritime University, Shanghai,  China.
  }}

 \date{}\maketitle

\begin{quotation}
 \textbf{Abstract}. Very recently,  Monz, et al. [arXiv:1507.08852] have reported the demonstration of factoring 15 using a scalable Shor algorithm with an ion-trap quantum computer. In this note, we remark that the report is somewhat misleading  because there are three flaws in the proposed circuit diagram of Shor algorithm.  We also remark that  the principles behind the demonstration have not been explained properly, including its correctness and complexity.

 \end{quotation}

 \section{Introduction}
It is well known that factoring an integer $n$ can be reduced to finding
the order of some integer $x$  modulo  $n$, i.e.,  $\mbox{ord}_n(x).$
So far, there is not a polynomial time algorithm run on classical computers which can be used to compute $\mbox{ord}_n(x)$.
In 1994,  Shor \cite{S97} proposed  an algorithm running on quantum computers which can compute $\mbox{ord}_n(x)$ and was claimed to be polynomial. The Shor  algorithm  requires two quantum
registers. At the beginning of the algorithm, one has to
  find $q=2^s$ for some integer $s$ such that $n^2 \leq  q < 2n^2$, where $n$ is to be factored. It then proceeds as follows.

  \begin{itemize}\parskip -2mm
\item{} \textit{Initialization}. Put register-1 in the uniform superposition
$ \frac 1{\sqrt q}\sum_{a=0}^{q-1}|a\rangle|0\rangle. $

\item{} \textit{Quantum modular exponentiation}. Keep $a$ in
  register-1 and
  compute $x^a$ in register-2 for some randomly chosen integer $x$.  It gives the state
$ \frac 1{\sqrt q}\sum_{a=0}^{q-1}
|a\rangle|x^a\rangle. $

\item{} \textit{Quantum Fourier Transformation}.   Performing Fourier transform on register-1, we obtain the state
$ \frac 1{q}\sum_{a=0}^{q-1}\sum_{c=0}^{q-1}
\mbox{exp}(2\pi iac/q)|c\rangle|x^a\rangle.
$

\item{} \textit{Observation}.  It suffices  to
observe register-1.  The probability $p$ that the machine reaches the  state $|c, x^k\rangle$ is
$$  \left| \frac{1}{q}\, \sum_{a:\, x^a \equiv x^k} \mbox{exp}(2\pi i ac/q) \right|^2, $$
where $ 0\leq k< r=\mbox{ord}_n(x) $,  the sum is over all $a\, (0 \leq a < q)$  such that $x^a\equiv
x^k$.

\item{} \textit{Continued Fraction Expansion}.
If there is a $d$ such that $\frac{-r}2 \leq dq-rc\leq \frac r 2  $,
then the probability of seeing $|c, x^k\rangle$ is greater than $1/3r^2$.
Since $q\geq  n^2$, we have
$$ \left|\frac c q-\frac d r \right|\leq \frac 1 {2q}\leq  \frac 1 {2n^2} <  \frac 1 {2r^2}.$$
 Then  $d/r$ can be obtained by rounding $c/q$.
\end{itemize}

Since 2001, some teams \cite{VS01,LW07,LB07,L12}
 have reported that they had successfully factored 15 into $3\times 5$ using Shor algorithm. We \cite{CCL14} have argued that all  these demonstrations are false and misleading, because they violate the necessary condition that the selected number $q$ must satisfy $n^2 \leq  q < 2n^2$. Otherwise, one can not complete the step of Continued Fraction Expansion in Shor algorithm.
 Essentially, these demonstrations have no relation to the Shor algorithm.

In 2015,  Monz, et al. \cite{M15} have reported a new  demonstration of factoring 15 using a scalable Shor algorithm with an ion-trap quantum computer. In this note, we want to remark that the report is misleading  because there are three flaws in the proposed circuit diagram of Shor algorithm. The claim  \cite{M15} that the scalable Shor algorithm is based on the Kitaev's result \cite{K95} is false.  In fact, Kitaev only claimed that factoring can be reduced to the Abelian Stabilizer Problem (ASP). He did not claim  that $2\ell$ qubits subject to a Quantum Fourier Transformation  (QFT) can be replaced by a single qubit  if only the classical information of the QFT (such as the period $r$) is of interest.
We also remark that the demonstration has no relation to Shor algorithm. Its principles have not been explained elaborately, including its correctness and complexity.

\section{The complexity of Shor algorithm}

 At the end of the  Shor  factoring  algorithm,
one should observe the first register and denote the measured result as an integer $c$. Its complexity argument comprises:
\begin{itemize}
\item[(1)]
The probability $p$ of seeing a quantum state $ |c, x^k \,(\mbox{mod}\, n)\rangle$ such that $r/2
\geq \{rc\}_q$ is greater than $ 1/{3r^2}$, where $n$ is the integer to be factored, $q$ is a power of 2 satisfying $n^2 \leq  q < 2n^2$
and $r=\mbox{ord}_n(x)$. For convenience, the notation $\mbox{mod}\, n$ will be omitted henceforth.
\item[(2)] There are $\phi(r)$ possible $c$ which can be used to compute the order $r$.
\item[(3)] The measured number in the second register, i.e., $x^k$,  takes  $r$
possible values $1, x, x^2, \cdots, x^{r-1}$.
\item[(4)] The success
probability of running the algorithm once is greater than  $r\cdot \phi(r)\cdot  \frac 1{3r^2}$.
By  $\phi(r)/r> \xi/\log\log r$ for some constant $\xi$, it concludes that the algorithm runs in polynomial time.
\end{itemize}

The Shor complexity argument can be depicted by the following Graph-1.

\hspace*{8mm}\begin{minipage}{\linewidth}
\includegraphics[angle=0,height=6.5cm,width=12cm]{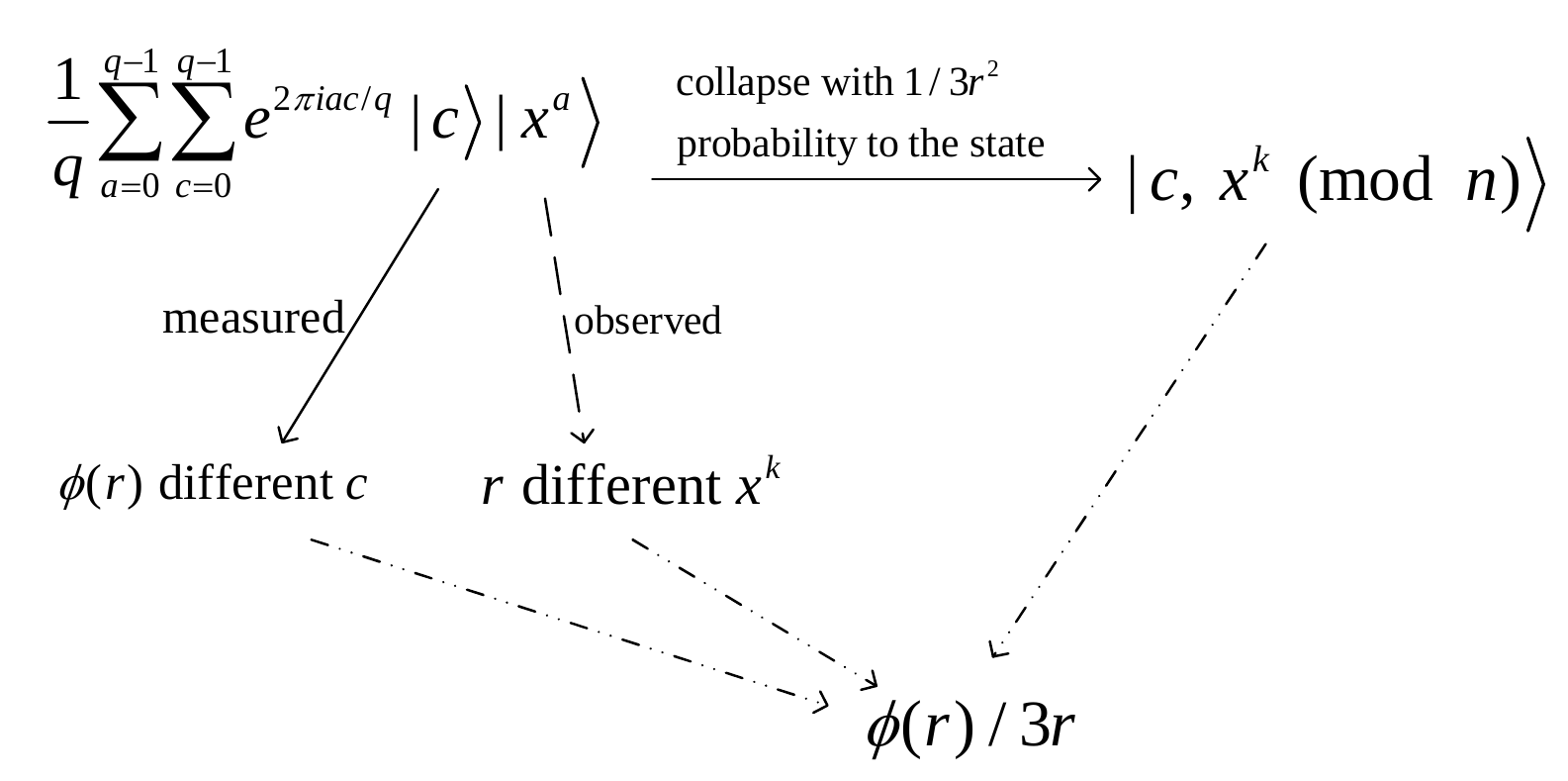}
 \end{minipage}

\centerline{\textsf{Graph-1: Shor complexity argument}}

\section{Review of the demonstration of Shor algorithm}

Very recently, Monz et al. \cite{M15} have reported the demonstration of factoring 15 using a scalable Shor algorithm with an ion-trap quantum computer. See the following Figure 1 for the circuit diagram of the demonstration.

\hspace*{-3mm}\begin{minipage}{\linewidth}
\includegraphics[angle=0,height=10.5cm,width=16cm]{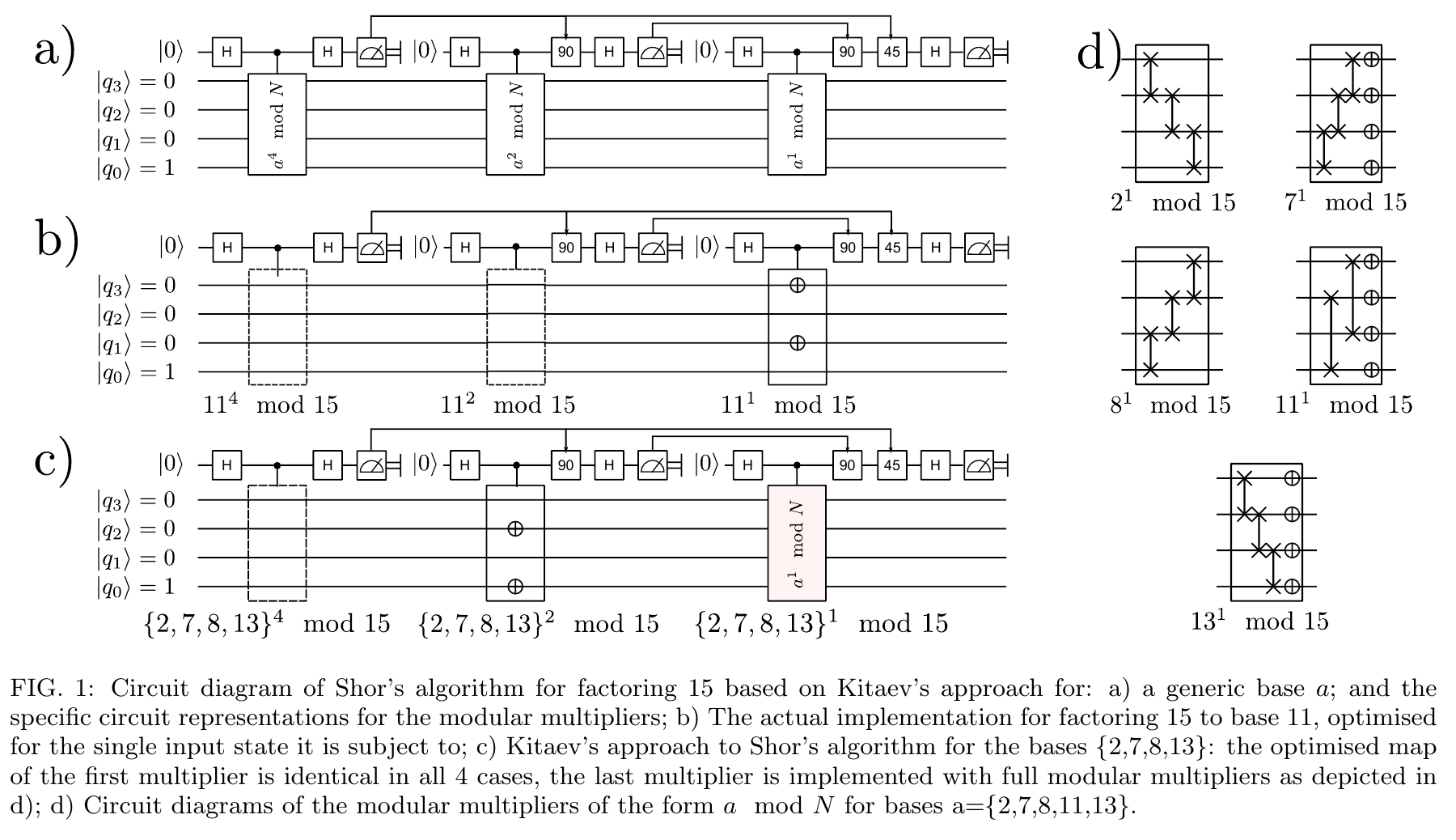}
 \end{minipage}

\section{Flaws in the demonstration of Shor algorithm}

As we see,  the original Shor algorithm requires at lest $3\ell$ qubits, where $\ell= \lceil\log_2(n)\rceil $, $n$ is to be factored. Among them,   $2\ell $ qubits are used in the register-1 and $\ell$ qubits are used in the register-2.
That is, the register-1 uses 8 qubits and the register-2 uses 4 qubits for factoring 15.

We find the demonstration of Shor algorithm does not satisfy the necessary condition. It only uses  one qubit for the register-1 and four  qubits for the register-2.  We here want to stress that
\begin{itemize}
\item[(1)] If less qubits are used in the register-1, then the step of Continued Fraction Expansion (CFT) in Shor algorithm can not be accomplished. In the new demonstration, the register-1 has only one qubit. Thus, the observed value $c$ is either $0$ or $1$. $q=2$. It directly violates the necessary condition for CFT.

    \item[(2)]  If less qubits are used in the register-1, then the Shor complexity argument fails because
    the probability $p$ of seeing a quantum state $ |c, x^k \,(\mbox{mod}\, n)\rangle$ such that $r/2
\geq \{rc\}_q$ is greater than $ 1/{3r^2}$  can not be properly estimated using the original Shor argument.

Here is a brief description of the original Shor complexity argument.  Setting $a=b r+k$ for some integer $b$ and the order $r=\mbox{ord}_n(x)$,  the probability $p$ is
$$\left|\frac 1{q}\sum_{b=0}^{\lfloor(q-k-1)/r \rfloor}e^{2\pi i(br+k)c /q}
 \right|^2.  $$
Then it argues that the probability $p$  equals to
$$\left|\frac 1{q}\sum_{b=0}^{\lfloor(q-k-1)/r \rfloor}e^{2\pi ib\{rc \}_q/q}
 \right|^2.  $$
Writing the above sum into an integral, we
obtain $$ \frac 1{q}
\int_{b=0}^{\lfloor\frac{(q-k-1)}{r} \rfloor}e^{2\pi
ib\{rc\}_q/q}\mbox{d} b \\
+ O\left( \frac{\lfloor (q-k-1)/r \rfloor}{q}(e^{2\pi i
\{rc\}_q/q}-1)\right).
 $$
  Taking $u =
rb/q$, we have
$$\frac {1}{r} \int_0^{\frac{r}{q} \lfloor \frac{q-k-1}{r}\rfloor} \mbox{exp}
\left(2\pi i u \frac{\{rc\}_q}{r} \right)\mbox{d}u $$ Taking into account that $k < r$,
  we can obtain the approximation $$\frac {1}{r} \int_0^{1} \mbox{exp}
\left(2\pi i u \frac{\{rc\}_q}{r}  \right)\mbox{d} u. $$
Hence, we have
$$ \left|\frac {1}{r} \int_0^{1} \mbox{exp}
\left(2\pi i u \frac{\{rc \}_q}{r} \right)\mbox{d} u \right|\geq 2/(\pi r)$$
 Therefore,
  $$p \geq \frac{4}{\pi^2 r^2}>1/3r^2.$$

\item[(3)] To modulate the wanted quantum state in the register-2, one has to find an efficient and universal quantum modular exponentiation method, which does not depend on any special bases. But the proposed circuit diagram shows that the method does depend on the different bases. It is not universal. Concretely, for base 11 one has to use the circuit b). For bases 2, 7, 8, 13, the circuit c) should be used. But even worse, both two circuits cannot generate the wanted quantum state in the register-2 in the original Shor's algorithm.

\end{itemize}

\section{Further discussions}

It  claims  \cite{M15} that the scalable Shor algorithm is based on the Kitaev's result \cite{K95}.
In the  page 2 (left column) of \cite{M15}, it writes:
\begin{quotation}
\noindent In Ref. 9 Kitaev
notes that, if only the classical information of the
QFT (such as the period $r$) is of interest, $2n$ qubits subject
to a QFT can be replaced by a single qubit. This
approach, however, requires qubit-recycling (specifically:
in-sequence single-qubit readout and state reinitialization)
paired with feed-forward to compensate for the reduced
system size. \end{quotation}

  But we find the claim is wrong. In fact, Kitaev only claimed that factoring can be reduced to the Abelian Stabilizer Problem (ASP). Throughout the paper \cite{K95},  He did not claim  that $2\ell$ qubits subject to a Quantum Fourier Transformation  (QFT) can be replaced by a single qubit  if only the classical information of the QFT (such as the period $r$) is of interest.

 We here  want to stress that the authors \cite{M15} misunderstand the purpose of using QFT in Shor algorithm.  The aim of applying QFT to the register-1 is to accumulate the wanted state $|c, x^k \,(\mbox{mod}\, n)\rangle$  such that it is possibly observed with significant probability. QFT has no direct relation to the order $r$. In fact, $r$ is deduced by rounding $\frac{c}{q}$.

Finally, we would like to stress that if the implementation is indeed credible then it would be a new quantum factoring algorithm, not the Shor algorithm, because all requirements of the original Shor's algorithm are not satisfied.

\section{Conclusion}

We remark that the demonstration of quantum factorization  proposed by Monz et al.  does not conform to the Shor algorithm. The necessary step of Continued Fraction Expansion in Shor algorithm can not be accomplished. The original Shor complexity argument does not apply to the demonstration. It does not specify a universal quantum modular exponentiation method. In view of these flaws, we do not think that the demonstration is believable. Naturally speaking, it has no relation to the Shor algorithm.


\begin{thebibliography}{4}

\renewcommand{\baselinestretch}{1.0}
  \renewcommand{\arraystretch}{.9}
  \normalsize \small \parskip 1mm


\bibitem{S97}  P. Shor: Polynomial-time algorithms for prime factorization and discrete
logarithms on a quantum computer. SIAM J. Comput.  26 (5): 1484-1509 (1997)


\bibitem{VS01}  L. Vandersypen, et al.: Experimental Realization of Shor Quantum Factoring Algorithm Using Nuclear Magnetic Resonance, Nature 414 (6866): 883-887, arXiv:quant-ph/0112176 (2001)

\bibitem{LW07}  B. Lanyon, et al.: Experimental Demonstration of a Compiled Version of Shor Algorithm with Quantum Entanglement", Physical Review Letters 99 (25): 250505. arXiv:0705.1398 (2007)

\bibitem{LB07}  C.Y. Lu,  et al.: Demonstration of a Compiled Version of Shor Quantum Factoring Algorithm Using Photonic Qubits, Physical Review Letters 99 (25): 250504, arXiv:0705.1684 (2007)

\bibitem{L12} E. Lucero, et al.: Computing Prime Fctors with a Josephson Phase Qubit Quantum Processor. Nature Physics 8, 719-723, 2012. arXiv:1202.5707 (2012)

\bibitem{CCL14} Z.J. Cao,  Z.F. Cao and L.H. Liu:  Remarks on Quantum Modular Exponentiation and Some Experimental Demonstrations of Shor Algorithm,  \verb"http://arxiv.org/abs/1408.6252" (2014)

 \bibitem{M15}  T. Monz, et al.: Realization of a scalable Shor algorithm.  \verb"http://arxiv.org/abs/1507.08852" (2015)

  \bibitem{K95} A. Kitaev:  Quantum measurements and the Abelian Stabilizer Problem. \\ \verb"http://arxiv.org/abs/quant-ph/9511026" (1995)

\end{thebibliography}
\end{document}